\documentclass[12pt]{article}
\usepackage{epsfig}

\def\beq{\begin{equation}}
\def\eeq#1{\label{#1}\end{equation}}
\def\eeqn{\end{equation}}

\def\beqa{\begin{eqnarray}}
\def\eeqa#1{\label{#1}\end{eqnarray}}
\def\eeqan{\end{eqnarray}}

\let\bar=\overbar

\def\Dslash{\not{\hbox{\kern-4pt $D$}}}
\def\dslash{\not{\hbox{\kern-2pt $\del$}}}

\def\msb{{\bar{\ssstyle M \kern -1pt S}}}

\usepackage{fancyhdr,graphicx}
\def\Title#1{\begin{center} {\Large {\bf #1} } \end{center}}

\begin{document}

\Title{Evolving Neutron Star Low-Mass X-ray Binaries to 
Ulta-compact X-ray Binaries}

\bigskip\bigskip


\begin{raggedright}

{\it Xiang-Dong Li\index{Li, X.-D.}\\
Department of Astronomy\\
Nanjing University\\
Nanjing 210093\\
P. R. China\\
{\tt Email: lixd@nju.edu.cn}}
\bigskip\bigskip
\end{raggedright}

\section{Introduction}

Mass transfer in neutron star low-mass X-ray binaries (NS LMXBs) is 
either driven by loss of orbital angular momentum or nuclear evolution
of the donor star, causing the orbit to the shrink or expand respectively.
The  ``bifurcation
period" $P_{\rm bif}$, the initial binary orbital period which
separates the formation of converging systems  from
the diverging systems (Tutukov et al. 1985)
was found to be  in the range $\sim 0.4-0.7$ day for LMXBs, and strongly 
dependent on magnetic braking (MB) (Pylyser \& Savonije 1988). 
For sufficiently small initial orbital periods, the evolution may
lead to the formation of of ultra-compact X-ray binaries (UCXBs)
with orbital periods ($P<50$ min), in which the donor is  a
white dwarf or a compact core of an evolved giant star
(Nelson et al. 1986; Tutukov et al. 1987; Pylyser \& Savonije
1988; Podsiadlowski et al. 2002; van der Sluys et al. 2005).

In this work we present the results on the evolution of NS LMXBs
and the formation of UCXBs (Ma \& Li 2009 for details). We consider the 
following processes related
to mass and angular momentum
loss mechanisms in LMXB evolution. (1) The standard MB law
(Verbunt \& Zwaan 1981; Rappaport et al. 1983) was shown to be 
contradicted with the 
observation of young stars in open clusters, and a modified
version was proposed (Sills et al. 2000; Andronov et al. 2003). (2) There is strong
evidence that during LMXB evolution the mass transfer is highly
non-conservative. Possible
ways of mass loss include ``evaporation" of
the donor  (Ruderman et al. 1989) or ``radio-ejection" of the transferred
material (Burderi et al.  2001, 2002; D'Antona et al. 2006) due to the pulsar
radiation/wind impinging on. In the latter case, the matter is lost
from the system at the inner Lagrangian ($L_1$) point, carrying away
angular momentum and altering the period evolution. Additionally,
a small fraction of the mass lost
from the donor may form a circum-binary (CB) disk around the binary
rather accretes onto the NS  (van den Heuvel 1994).

\section{Evolution code and binary mode}

\subsection{The stellar evolution code}
We use an updated version of the stellar evolution code originally
developed by (Eggleton 1971, 1972) to calculate the evolutions 
of binaries
consisting of an NS (of mass $M_1$) and an MS secondary (of mass
$M_2$). For the secondary star we assume a solar chemical
composition ($X=0.70$, $Y=0.28$, and $Z=0.02$). We
assume that the spin of the secondary star and the binary
orbital revolution are always synchronized. Assuming rigid body
rotation of the secondary star and neglecting the spin angular momentum
of the neutron star, the total angular momentum $J$ of the binary system
can be expressed as
\begin{eqnarray}
J&=&I_2\omega+J_{\rm orb} \nonumber \\
&=&I_2\omega+G^{2/3}M_1M_2(M_1+M_2)^{-1/3}\omega^{-1/3}
\end{eqnarray}
where $I_2$ is the moment of inertia of the secondary star, 
$G$ the gravitational constant, and $\omega$
the angular velocity of the binary.

We consider three kinds of mechanisms of angular momentum loss. The
first is the angular momentum loss due to gravitational radiation
\begin{equation}
\frac{{\rm d}J_{\rm{GR}}}{{\rm d}t}=-\frac{32}{5}\frac{G^{7/2}}{c^5}
\frac{M_1^2M_2^2(M_1+M_2)^{1/2}}{a^{7/2}},
\end{equation}
where $a$ is the orbital separation and $c$ is the speed of light.

The second angular momentum loss mechanism is for non-conservative
mass transfer. We
assume that a small fraction $\delta (\ll
1)$ of the mass lost from the donor feeds into the CB disk rather
leaves the binary, which yields a mass injection rate of the CB disk
as $\dot{M}_{\rm CB}=-\delta\dot{M_2}$. Tidal torques are then
exerted on the binary by the CB disk via gravitational interaction,
thus extracting the angular momentum from the binary system. The
angular momentum loss rate via the CB disk is estimated to be
(Taam \& Spruit 2001)
\begin{equation} \label{cb}
\frac{{\rm d}J}{{\rm d} t}|_{\rm CB}=-\gamma\left(\frac{2\pi
a^2}{P}\right) \dot{M}_{\rm CB}\left(\frac{t}{t_{\rm
vi}}\right)^{1/3},
\end{equation}
where $\gamma^{2}=r_{\rm i}/a=1.7$ ($r_{\rm i}$ is the inner radius
of the CB disk), $t$ is the time since mass transfer begins. In the
standard $\alpha$-viscosity disk (Shakura \& Sunyaev 1973), the viscous
timescale $t_{\rm vi}$ at the inner edge $r_{\rm i}$ of the CB disk
is given by $ t_{\rm vi}=2\gamma^{3}P/3\pi\alpha\beta^{2}$, where
$\alpha$ is the viscosity parameter (we set $\alpha=0.01$ in the
following calculations), $\beta=H_{\rm i}/r_{\rm i}\sim 0.03$
(Belle et al. 2004), and $H_{\rm i}$ is the scale height of the disk.
We also assume that
the NS accretion rate is limited to the Eddington accretion rate,
and that when the mass transfer rate is less than $\dot{M}_{\rm
Edd}$,  half of the mass is accreted by the NS, i.e., $\dot{M}_1=
\min (\dot{M}_{\rm Edd}, -\dot{M}_2/2)$. The excess mass is
lost in the vicinity of the NS through isotropic winds, carrying
away the specific angular momentum of the NS, i.e.
\begin{equation} \label{eq:massloss}
 \frac{{\rm d} J}{{\rm d}t}|_{\rm ML} \simeq  \left\{
 \begin{array}{lll}
\frac{1}{2}\dot{M}_2 a_{1}^2\omega, & & |\dot{M}_2| < 2\dot{M}_{\rm Edd}\\
(\dot{M}_2+\dot{M}_{\rm Edd}) a_{1}^2\omega, & & |\dot{M}_2 |\ge 2\dot{M}_{\rm Edd}
\end{array}\right.
\end{equation}
where $a_1=aM_2/(M_1+M_2)$ is the orbital radius of the NS,  and
$\omega$ is the orbital angular  velocity of the binary.

The third angular momentum loss mechanism is MB. We use the saturated magnetic
braking law suggested in  Sills et al. (2000),
\begin{equation} \label{eq:magneticbraking}
 \frac{{\rm d} J}{{\rm d}t}|_{\rm MB} =  \left\{
 \begin{array}{lll}
-K \omega^3 \left(\frac{R_2}{R_{\odot}}\right)^{1/2}
\left(\frac{M_2}{M_{\odot}}\right)^{-1/2},  & & \omega \leq \omega_{\rm{cr}} \\
 -K \omega_{\rm{cr}}^2 \omega
\left(\frac{R_2}{R_{\odot}}\right)^{1/2}
\left(\frac{M_2}{M_{\odot}}\right)^{-1/2}, & & \omega >
\omega_{\rm{cr}}
\end{array}\right.
\end{equation}
where $K=2.7\times10^{47}$ gcm$^2$s  (Andronov et al. 2003),
$\omega_{\rm{cr}}$ is the critical angular velocity at which the
angular momentum loss rate reaches a saturated state, and can be
estimated as ( (Krishnamurthi et al. 1997),
\begin{equation}
\omega_{\rm{cr}}(t) = \omega_{\rm{cr},\odot}
\frac{\tau_{\rm{t}_0,\odot}}{\tau_{\rm{t}}},
\end{equation}
where $\omega_{\rm{cr},\odot}=2.9\times 10^{-5}$ Hz,
$\tau_{\rm{t}_0,\odot}$ is the global turnover timescale for the
convective envelope of the Sun at its current age, $\tau_{\rm{t}}$
for the secondary at age $t$, solved by integrating the inverse
local convective velocity over the entire surface convective
envelope (Kim \& Demarque 1996). 

\begin{figure}[htb]
\begin{center}
\epsfig{file=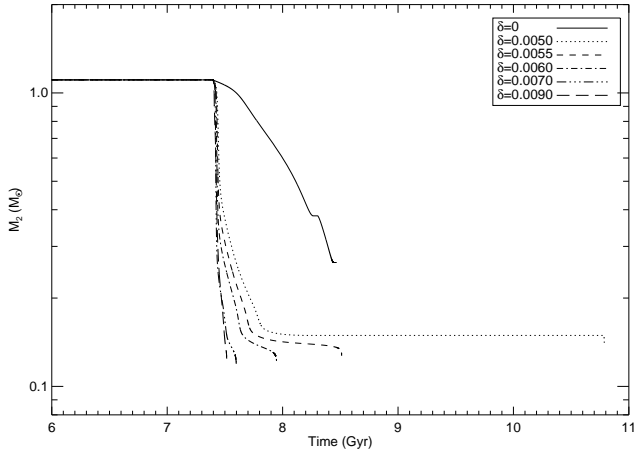,height=2.5in}
\caption{Evolution of the donor mass for different values of 
the CB disk parameter $\delta$.}
\label{}
\end{center}
\end{figure}
\begin{figure}[htb]
\begin{center}
\epsfig{file=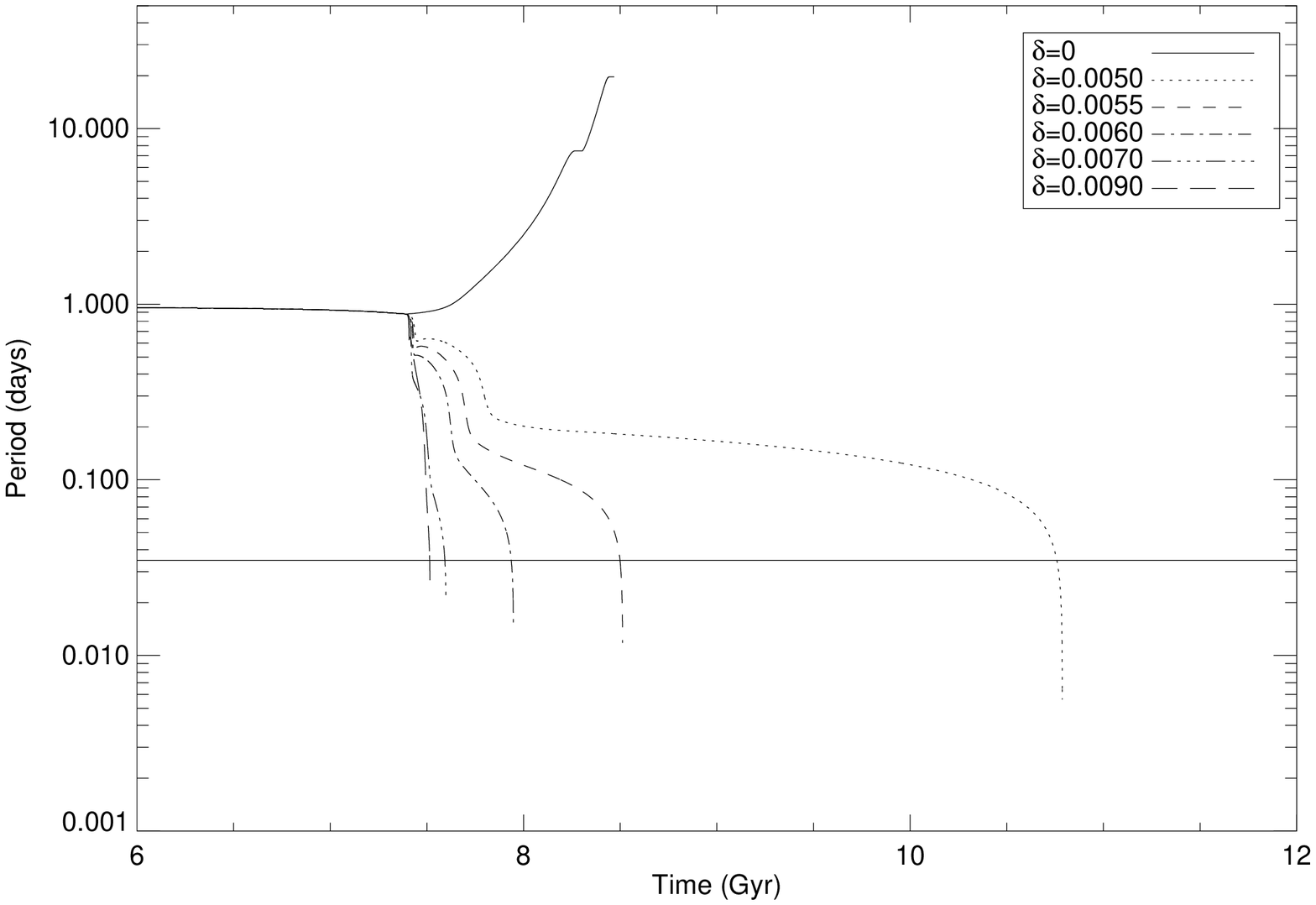,height=2.5in}
\caption{Evolution of the orbital period for different values of 
the CB disk parameter $\delta$.}
\label{}
\end{center}
\end{figure}

\section{Results}

The formation and evolutionary paths of UCXBs depend on the adopted
values of $\delta$. To illustrate the effects
of $\delta$ on the binary evolution, in Figs.~1 and 2,
we plot the evolution of  the donor mass and period as a function of
age respectively, for a binary system with
$M_{2,\rm i}=1.1M_\odot$, $P_{\rm i}=1.04$ d and different values of
$\delta$. 
A larger value of $\delta$ leads to shorter formation time, as seen from
Fig.~2; if $\delta$ is too small, the binary will not be able
to reach the $50$ min period within $13.7$ Gyr due to inefficient
angular momentum loss. 
When $\delta<0.0055$, the orbital
period first decreases with mass transfer until the donor star loses
its outer envelope and shrinks rapidly at $P\sim 0.1-0.2$ d. This
causes a cessation of mass transfer. In the subsequent evolution the
orbital period may decrease down to the ultra-short regime under the
effect of GR, until the secondary star fills its RL again, and the
binary appears as a UCXB. When $\delta \ge 0.0055$ the binary
evolves directly into the ultra-short regime with decreasing orbital
period.

We need to mention that the distribution of $\delta$ depends on the
value of the viscous parameter $\alpha$. From Eqs.~(3) one can see that 
the CB disk-induced angular
momentum loss rate is proportional to
$\alpha^{1/3}\delta$. So if keeping $\alpha^{1/3}\delta$ 
constant, the binary evolution will be exactly the same.

\section{Comparison with observations}

There are currently 10 UCXBs with known periods, 5 of which are
persistent sources and 5 are transients. To compare
observations with our CB disk-assisted binary model, we plot the
$\dot{M}_1(=-\dot{M}_2/2)$ vs. $P_{\rm orb}$ relations in Fig.~\ref{f6} for
binary systems with $M_2=1.1M_\odot$, $P_{\rm i}=1.04$ d and
$\delta=0.005-0.009$. We also
indicate in Fig.~3 whether the accretion disks in
the LMXBs are thermally and viscously stable, according to the
stability criterion for a mixed-composition ($X=0.1$, $Y=0.9$) disk
from Lasota et al. (2008)
We use the symbols $\times$, $\ast$, and $+$ on the
evolutionary tracks to denote where the hydrogen composition $X$
of the donor becomes 0.3, 0.2, and 0.1, respectively.
The positions of UCXBs are marked in these two
figures with circles and triangles for persistent and transient sources,
respectively. Besides them, we also include 18 NS LMXBs
with known $P$ and $\dot{M}_1$ (data are taken from Liu et al. 2007; 
Watts et al. 2008; Heinke et al. 2009).

A comparison between our CB disk-assisted binary models and the
observations of (compact) NS LMXBs suggesting
that it is possible to form UCXBs from normal LMXBs. We note that
three of the UCXBs are in globular clusters, indicating low
metallicities in these systems. However, from our calculations we
find that change of metallicities does not significantly affect the
binary evolution when the CB disk is involved.

\begin{figure}[htb]
\begin{center}
\epsfig{file=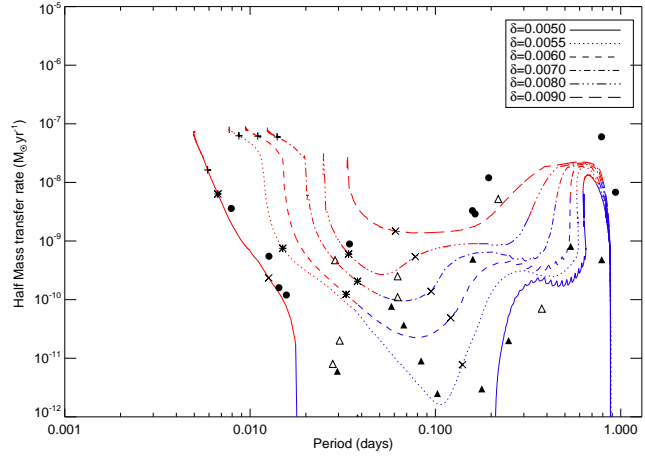,height=2.5in}
\caption{Evolution of the mass accretion rate vs. orbital period 
for different values of the CB disk parameter $\delta$.}
\label{}
\end{center}
\end{figure}



\section{Conclusion}
During mass transfer in LMXBs, a CB disk may be formed as a result of 
mass outflow from the
accretion disk, and has been invoked as an efficient process for
the removal of orbital angular momentum
 (Taam \& Spruit 2001). We propose a scenario for the formation of UCXBs
from L/IMXBs with the aid of a CB disk in this work. The suitable
binary parameter space ($M_{\rm 2,i}$ and $P_{\rm i}$) with
reasonable choice of the CB disk parameter $\delta$ for the
formation of UCXBs within $13.7$ Gyr is found to be significantly
larger than in previous ``magnetic capture" model
(van der Sluys et al. 2005). This difference is caused by the fact
that the bifurcation period is considerably increased if the CB disk
is included.

\bigskip
This work was
supported by Natural Science Foundation of China under grant
10873008 and National Basic Research Program of China (973 Program
2009CB824800).







\end{document}